# Two-Body Problem in General Relativity: A Heuristic Guide in Einstein's Work on the Einstein-Rosen Bridge and EPR Paradox

Galina Weinstein

November 18, 2015

*Between 1935 and 1936, Einstein was occupied with the Schwarzschild solution and the singularity within it while working in Princeton on the unified field theory and, with his assistant Nathan Rosen, on the theory of the Einstein-Rosen bridges. He was also occupied with quantum theory. He believed that quantum theory was an incomplete representation of real things. Together with Rosen and Boris Podolsky he invented the EPR paradox. In this paper I demonstrate that the two-body problem in general relativity was a heuristic guide in Einstein's and collaborators' 1935 work on the Einstein-Rosen bridge and EPR paradox.*

In 1921 Erich Trefftz constructed an exact static spherically symmetric solution (the de Sitter–Schwarzschild solution or the Weyl-Trefftz line element) for Einstein's vacuum field equations with the cosmological constant (Trefftz 1922). The Weyl-Trefftz metric represents a model for a spherical closed (finite) universe, a de Sitter static universe devoid of matter whose material mass is concentrated in just two spherical bodies on opposite sides of the world (Einstein 1922, 318, 448).

In a comment on Trefftz's paper, Einstein identified a problem with the Weyl-Trefftz line element. He demonstrated that the Weyl-Trefftz exact solution contained a true singularity in the empty space between the two bodies. In the empty space between the two masses the time variable vanishes. Consequently, time stands still in the de Sitter empty space between the two masses. This signifies there are other masses distributed between the two bodies.

Einstein said that Hermann Weyl had already shown that many masses existed somewhere in-between the two bodies (Einstein 1922, 449). Indeed, to keep the two bodies apart at a constant distance (in a static closed world), Weyl had to introduce a true singularity at the mass horizon (somewhere in the empty space between the two bodies), and to add a positive Einstein cosmological term to the field equations. He introduced a true singularity and concluded that a zone of matter exists between the two bodies. Weyl was misled by the apparent de Sitter singularity into believing that the mass in de Sitter's world is distributed on a mass horizon, and this induced him to introduce a true singularity.

Weyl's view, like Einstein's, was essentially the following: the de Sitter empty universe model solves Einstein's field equations with cosmological constant everywhere except on the equator.

On the equator the components of the metric tensor tend to zero as they do in the neighborhood of a material point (the Schwarzschild solution). Hence, the de Sitter world corresponds to a world the material content of which is concentrated in the equator.

Weyl therefore omitted part of the space-time around the horizon, and replaced it by the Schwarzschild interior solution of Einstein's field equations with the cosmological constant. This combined solution led him to the most general static, spherically symmetric solution of Einstein's vacuum field equations with cosmological constant, the Weyl-Trefftz line element. Hence, instead of removing the apparent singularity, Weyl introduced a true singularity by joining two solutions: the de Sitter and Schwarzschild interior solutions (Goenner 2001, 111-112; Weinstein 2015, 316-318).

In the same year 1921, Rudolf (Förster) Bach studied the de Sitter–Schwarzschild metric as a model for an equilibrium configuration consisting of two spherical bodies at rest. We obtain the Schwarzschild solution by taking the Bach-Weyl two-body solution and considering the limit as the mass of the second body tends to zero. Bach noted that this static solution required the presence of a singularity on the line connecting these two bodies, and that this fact violated the regularity of the solution. The bodies could not be in equilibrium under the influence of gravitational forces alone (Bach and Weyl 1922, 134-142).

Weyl published remarks as an addendum to Bach's paper, "The Static Two-body Problem", and focused on the interpretation of this type of singularity. He concluded that the physical significance of this interpretation should not be exaggerated. Consequently, there was no static solution for the real two-body problem (Bach and Weyl 1922, 145).

In 1935 Einstein was occupied with the Schwarzschild solution and the singularity within it while working in Princeton on the unified field theory and, with his assistant Nathan Rosen, on the theory of the Einstein-Rosen bridges (the Einstein-Rosen bridges paper).

In the Einstein-Rosen bridges paper of 1935, Einstein negated the possibility that particles were represented as singularities of the gravitational field because of his polemic with Ludwig Silberstein. Einstein said that Silberstein confirmed in his presence that one could not accept the possibility that material particles might be considered singularities of the metric field (Einstein and Rosen 1935, 73).

Silberstein thought he had demonstrated that general relativity was problematic. He constructed, for the vacuum field equations for the two-body problem, an exact static solution with two singularity points that lie on the line connecting these two points. The singularities were located at the positions of the mass centers of the two material bodies. Silberstein concluded that this solution was inadmissible physically and contradicted experience. According to his equations the two bodies in his solution were at rest and were not accelerated towards each other; these were non-allowed results and therefore Silberstein thought that Einstein's field equations should be modified together with his general theory of relativity.

Before submitting his results as a paper to the *Physical Review*, Silberstein communicated them to Einstein. This prompted Einstein's remark, in his paper with Rosen in 1935, that matter particles could not be represented as singularities in the field (Einstein and Rosen 1935, 73).

However, Einstein quickly changed his mind after the "bridges" paper. A year later, Einstein relegated this strict demand and was willing to admit singularities of this kind, and accepted singularities as representing material particles. In December 1935, Einstein objected to Silberstein's two-body solution, and found a lacuna in his solution claiming it had an additional singularity on the line connecting the two singularities (in Havas 1993, 106-107). Hence Silberstein's solution contained singularities to which Einstein objected in a final field theory. Since Silberstein's solution was found to be erroneous, Einstein realized that the bridge could not represent a particle and he returned to his previous idea that material particles could be considered as singularities of the field in general relativity.

We have seen that Einstein had given the very same explanation thirteen years earlier, in 1922. In 1921, Trefftz constructed an exact static spherically symmetric solution for Einstein's vacuum field equations with the cosmological term. The Trefftz metric represents a model for the two-body problem. In a comment on Trefftz's paper, Einstein identified a problem with the Trefftz line element. In 1922 Einstein demonstrated that the Trefftz exact solution contained a true singularity in the empty space between the two bodies (Einstein 1922, 448).

However, after the Einstein-Rosen bridges paper, Silberstein grew so satisfied with finding that Einstein mentioned his name that he insisted that the Einstein-Rosen bridge solution, a two-body problem, has singularities at A and B only, and not as in Bach's and Weyl's two-body solution, along the line joining the two bodies. He even noted that this has induced Einstein and Rosen to publish the Einstein-Rosen bridges paper. Silberstein became antagonistic to Einstein's general theory of relativity, criticized it, and entered into a controversial debate with Einstein. He sent the paper with his new idea to the *Physical Review* (Silberstein 1936) and informed Einstein about it (Weinstein 2015, 226-227).

In March 1936, Einstein wrote to Silberstein about his advice to him to withdraw publication. In addition, he said that the newspaper contained the idiotic claim that he had revised his general theory of relativity because of Silberstein's earlier letters to him. By his efforts Silberstein had made it necessary for Einstein (and Rosen) to correct his errors publicly in a *Physical Review* letter (Havas 1993, 107-113). And so on February 10, 1936 the Newspapers published the following announcement: [1]

> "Professor Albert Einstein, who promulgated his theory of relativity in 1905, has replied to criticism of the theory by Dr. Ludwik Silberstein of the University of Toronto in the current issue of the *Physical Review*.

---

[1] "Einstein Says Attack on His Work by Torontonian is Based on Error," *The Gazette, Montreal Monday*, February 10, 1936.

Dr. Einstein said Dr. Silberstein's conclusions were based on an error 'which the critic has not yet recognized'. […]

the work of L. Silberstein, relates to a two-centers' solution of the gravitational field equations. The author believes that the result of his simple calculation contradicts the general theory of relativity, since he sees in this result an elementary contradiction to the facts of gravitation.

'I have already informed Mr. Silberstein that his result is based on an error, which, unfortunately, he has so far failed to realize […]'".

Einstein was asked to:

"State his attitude on reports in some newspapers that he was working on a new theory of matter as a result of the 'defects' in his original theory brought to his attention by Dr. Silberstein. Dr. Einstein, in collaboration with Dr. N. Rosen of the Institute of Advanced Studies in Princeton, N.J., published his paper outlining his new theory of matter, linking the theory of relativity and the quantum theory, in the *Physical Review* issue of July 4, 1935. Dr. Einstein's statement makes it clear that Dr. Silberstein's mathematical efforts had nothing to do with the formulating by Dr. Einstein and Dr. Rosen of their new theory of matter, upon which, it has been known, Dr. Einstein has been working for many years".

In response Silberstein wrote to "Sweet Mr. Einstein", and the Silberstein solution was never mentioned again (Einstein and Rosen 1936).

The singularities haunted Einstein; the first singularity that seemed to trouble him was the Schwarzschild singularity. Einstein and Rosen were trying to permanently dismiss the Schwarzschild singularity and adhere to the fundamental principle that singularities of the field are to be excluded. Einstein explained that one of the imperfections of the general theory of relativity was that as a field theory it was not complete in the following sense: it represented the motion of particles by the geodesic equation. A mass point moves on a geodesic line under the influence of a gravitational field. However, a complete field theory implements only fields and not the concepts of particle and motion. These must not exist independently of the field but must be treated as part of it. That is the reason why material particles could not be considered singularities of the field. Einstein wanted to demonstrate that the field equations for empty space are sufficient to determine the motion of mass points.

In 1935 Einstein attempted to present a satisfactory treatment that accomplishes a unification of gravitation and electromagnetism. For this unification, or as he called it the combined problem, he needed a description of a particle without singularity. In 1935, Einstein and Rosen joined two Schwarzschild solutions at the Schwarzschild limit and omitted part of the space-time beyond the Schwarzschild singularity. They showed that it was possible to do this in a natural way and they proposed the bridges solution (Einstein and Rosen 1935, 76): The four dimensional space is described mathematically by two congruent sheets that are joined by a hyperplane in which the metric tensor vanishes and the determinant of the metric tensor vanishes. Einstein and Rosen called such a connection between the two sheets a "bridge". The bridge, which connects the two sheets, is spatially finite and characterizes or describes the presence of an electrically neutral elementary particle. With this conception Einstein thought he could represent an elementary particle by using only the field equations. The two sheets can also be interpreted as each corresponding to the same physical space (Einstein 1936, 379-380).

Einstein and Rosen added an electromagnetic field to their solution, so that it could also represent a charged particle. If several particles are present, this case corresponds to several bridges. Einstein did not succeed in finding a multi-bridge solution of his modified field equations, and could not describe the whole field without introducing singularities. And indeed, Einstein soon decided to abandon the bridge as a non-singular particle model.

The two-body problem in general relativity (unified field theory) served as a model for Einstein and Rosen when in 1935 they worked with Boris Podolsky on the Einstein–Podolsky–Rosen (EPR) argument (Einstein, Podolsky and Rosen 1935).

Recall that the singularities haunted Einstein. Einstein and Rosen were trying to permanently dismiss the Schwarzschild singularity and adhere to the fundamental principle that singularities of the field are to be excluded. Einstein explained that one of the imperfections of the general theory of relativity was that as a field theory *it was not complete* because it represented the motion of particles by the geodesic equation. Einstein also searched for "complete descriptions of physical conditions" in quantum mechanics (Einstein 1936, 375).

The two-body problem in general relativity was a heuristic guide in the search of a solution to the problem that the ψ function cannot be interpreted as a complete description of a physical condition of one system (Weinstein 2015, 238). Einstein explained this argument in his 1936 paper, "Physics and Reality" (Einstein 1936, 376). Consider a mechanical system comprising two partial systems A and B that are in interaction with each other only during a limited time, and let there be given the wavefunction ψ before their interaction. After the interaction between them, the system A is separated from B. The Schrödinger equation supplies the ψ function for the two partial systems A and B after the interaction has taken place. Let us now determine the physical condition of the partial system A as completely as possible by measurements. Quantum mechanics then allows us to determine the ψ function of the partial system B from the measurements made and from the ψ function of the total system. This determination, however, gives a result that depends upon which one of the values specifying the physical condition of A has been measured (for instance coordinates or momenta). Einstein could not accept this situation and claimed that the ψ function could therefore not be interpreted as a complete description of the physical condition of the system.